\documentclass[%
 aip,
 jmp,%
 amsmath,amssymb,
 reprint,%
]{revtex4-1}

\usepackage{graphicx}
\usepackage{dcolumn}
\usepackage{bm}

\begin{document}

\preprint{AIP/123-QED}

\title{Logistic Growth for  the Nuzi Cuneiform Tablets:\\ 
Analyzing Family Networks  in 
Ancient Mesopotamia
}

\author{Sumie Ueda}
\affiliation{The Institute of Statistical Mathematics,
10-3 Midori-cho, Tachikawa, Tokyo 190-8562, Japan%
}%

\author{Kumi Makino}%
\affiliation{ Kamakura Women's University, 6-1-3 Ofuna  Kamakura
 Kanagawa 247-8512 Japan
}%

\author{Yoshiaki Itoh}
\affiliation{The Institute of Statistical Mathematics,
10-3 Midori-cho, Tachikawa, Tokyo 190-8562, Japan%
}%
\author{Takashi Tsuchiya }
\affiliation{National Graduate Institute for Policy Studies, 7-22-1 Roppongi 
 Minatoku Tokyo 106-8677 Japan%
}%


\begin{abstract}

We  reconstruct  the year   of publication of  each cuneiform tablet  
of the Nuzi society in ancient Mesopotamia. 
The  tablets, are  
on land transaction, marriage, loan,  slavery contracts  etc. 
The number of tablets seem to increase by  
logistic growth  until saturation.  It  may  show      the dynamics of 
concentration of lands or  other
 properties 
 into 
 few powerful  families  in a  period of about twenty years.   
We reconstruct  family trees and social  networks of Nuzi 
and  estimate 
the publication years of cuneiform tablets consistently with the trees and networks, 
  formulating  least squares problems with
linear inequality constraints. 
  
\end{abstract}

\maketitle
The observed phase changes of phenomena in 
human sciences are sometimes well explained by using the models developed in physics. 
The  neolithic transition (i.e., the shift from hunter-gatherer to
agricultural economies) \cite{fm},   change of word ordering rules in natural  
languages \cite{iu2},  
change of election systems  in the present  society \cite{iu}, and many 
interesting problems. 

 The time evolution of Networks based
on human interactions in economy,  politics,   transportation systems,  
Internet,  family trees, etc.  
 is drawing the attention to the present physicists \cite{krl, ims}.  
 Here we make use of the interactions in the network of family
 trees to estimate the years   of the contracts given in the Nuzi cuneiform tablets 
in ancient 
Mesopotamia to find the simplest nonlinear dynamics,  logistic growth of the number of 
cuneiform tablets.

\begin{figure}[t]
\special{epsfile=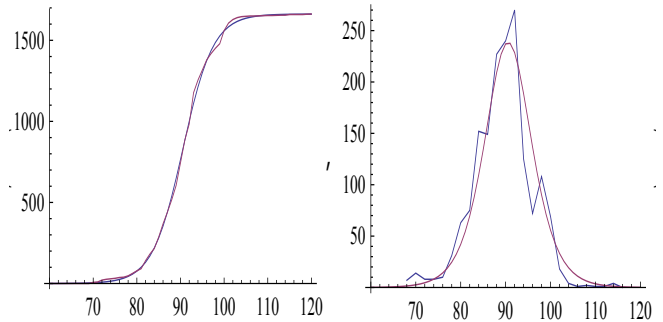 hscale=0.9  vscale=1.5} 
\vspace{5.0cm}
\caption{ Data (publication years   obtained by mathematical optimization) and logistic growth. 
 Left figure: data and estimated logistic 
distribution function. Right figure: data and estimated logistic probability density.  
The value of the maximum  log likelihood for the logistic 
distribution is 
-5404.04, while that of normal distribution is -5438.27.}\label{logistic}
\end{figure}

The Nuzi society   
\cite{ri, ch, mo}, 
existed  for  five or six generations in  about  15th century B.C.
`Nuzi Personal Names'  \cite{gpm} (NPN) 
is   the index for  personal names with 
the kinships for the  
cuneiform  tablets.  
We  introduce the time coordinate of publication of  each cuneiform tablet, 
as well as    the birth year and the death year  
of each person 
 listed  in NPN,  Fig \ref{logistic}. 
We  formulating  least squares problems with
linear inequality constraints making use of 
conditions  obtained by kinship relations and other information 
in  NPN.  
For example we 
assume that 
a father is at least 15 years older than 
his son, contractors were living at the time of the contract, etc. 
 It seems that  publication  of the cuneiform tablets (documents)  are concentrated on 
a 
period  of twenty years and that they 
   continued to be published 
  until saturation  by 
logistic growth   Fig \ref{logistic}. 
The large  portion of the 
documents are on land transaction,  for example by false adoption given
 later JEN 208 \cite{ch}.  
The logistic 
 growth   seems to show    the dynamics of 
 the emergence of 
social hierarchy, which is 
  well recognized  in philological studies    
\cite{ja, le}, namely 
the increase   of concentration of lands or  other
 properties 
 into 
 few powerful  families \cite{u}, as the Te\b{h}iptilla family \cite{mai, mai2}. 
 The  logistic growth seems to be  
 natural to understand  the accumulation   of cuneiform tablets in Nuzi and  support to 
convince  
the success of our numerical study.  
Our  time obtained from the least square  will reconstruct the life of  Te\b{h}iptilla and other 
persons in Nuzi in ancient Mesopotamia.


   Nuzi is known for its unique false adoption documents as shown below 
in Example 1, the cuneiform document JEN 208.
The adopted members protected the 
adoptive father's  family in exchange for their properties, such as land estates.  
There exists a family called Te\b{h}iptilla's family, and this family was
involved in many contracts and transactions. 
  {JEN 208}.   
 is 
the contract between  Ilu\b{i}a,  a son of \b{H}amattar  and Te\b{h}iptilla a son of
 Pu\b{h}i\v{s}enni, with the names of witnesses and scribes.  The documents 
 JEN are given in  \cite{ch},  and  JENu, given later,   are unpublished Nuzi tablets 
excavated by the Iraq Museum and the American Schools of Oriental Research in 1926.  

\noindent
\noindent
{\bf Example 1.} (\cite{ch,ma})
{\bf JEN 208} 
Adoption tablet.  Ilu\b{i}a, a son of \b{H}amattar. He made, Te\b{h}iptilla  
 son of  Pu\b{h}i\v{s}enni, for sonship. 
Ilu\b{i}a assigned Te\b{h}iptilla to  as (his) share 2 imer and 3 awehari of land 
in the large standard, 
west of the dimtu of Imbi-ili-su, east of the dimtu of eniia. And Te\b{h}iptilla  [ga]ve to 
Ilu\b{i}a
as his gift 10 imer of barley. If the land ge[ts] a claimant,   Ilu\b{i}a shall clear (it), 
to Te[\b{h}ip-]tilla  he (shall) restore (it). 
The ilku service of the land only [I]lu\b{i}a [shall] bear/ [If Ilu\b{i}a infr]inges (the agreement), 
[he shall furnish 1 mina of silver (and) 1 mina of gold]. (Rest of the obverse destroyed)
L1.17-20:4 seals; some seals destroyed.

[]:  broken

(): added by original translator

dimtu:  originally means a  tower, but also means a region surrounded by city wall, and
 a district. Jancowska pointed out that there were 71 dimtu in
 Nuzi \cite{ja2}
\vspace{0.3cm}

In NPN, the personal names originally written in cuneiform are included as the form of
 phonetic values with alphabet. For example a name Iluia in JEN 208 is expressed in 
three cuneiform characters and these phonetic values joined by hyphen as I-lu-ia.
 Cuneiform character sometimes has more than one phonetic value, and different 
values may be put by different readers or scholars such as I-lu-ya.
 We analyzed data only appears in NPN, we could  avoid this reading difference. 

Most names found in Nuzi documents are male and inscribed 
with his father,  as Te\b{h}iptilla, son of Pu\b{h}i\v{s}enni.  
 There are 124 persons named Ta\b{i}a, 7 persons named  Ilu\b{i}a, 1 
person named \b{H}amattar, etc. 
 We  take the sequence of phonetic values  listed at first in NPN.

In NPN, individual names are listed in alphabetical order with the information 
in the original cuneiform documents. Example 2 is the part of NPN associated
with two names ``Ilu\b{i}a'' and 
``\b{H}amattar'' who appear in JEN 208.  We see that there is one  \b{H}amattar. 
There are  7 Ilu\b{i}a and some of them represent the same person.

\noindent
{\bf Example 2.} Description on \b{H}amattar
and Ilu\b{i}a
 in Nuzi personal names
(A syllable is separated by $-$. The special character \b{h} is described by 
\verb+h_{u}+. The \v{s} is described by \verb+s^{v}+.)

\vspace{-1ex}

{\scriptsize
\begin{verbatim}
H_{u}AMATTAR
 H_{u}a-ma-at-ta-ar var. (2) H_{u}a-ma-at-ti-ir
  1) f. of I-lu-ia JEN 208:2; (2) JENu 414; gf. of Ta-a-a JEN
    369:4
ILUI_{n}A
 Ilu-ia var. (2) I-lu-ia
  1) s. of H_{u}a-ma-at-ta-ar (2) JEN 208:1 8 11 13 14; 369:3,
    10; H_{u}a-ma-at-ti-ir (2) JENu 414
  2) scribes. of ^{d}Sin-na-ap-s^{v}i-ir JEN 226:42 45; 438:21 24
  3) s. of U^{'}-zu-ur-me JEN 13:37
  4) f. of S^{v}a-ar-til-la JEN 640:13; 662:95; HSS IX 7:31 (read
    so against Ili-iddina of copy); 35:38; RA XXIII 33:33;
    50:43; 67:23
  5) f. of Ta-a-a JEN 369:3 101
  6) f. of Da-an-ni-mu-s^{v}a JEN 345:5
  7) scribe JENu 625; AASOR XVI 56:41
\end{verbatim}
}

These information include a represented name, the scribal variations,
 kinship relations, and the line number in which the name appears.
For example,  JEN 208:2 means that the name appears in the line 2 of the cuneiform
 tablet JEN 208.  In some case, only document is shown 
without a line number, 
like JENu 414.  
The contractor's names are usually included in the first 10 lines, about 2 to 5 lines, in each documents, while 
the names of witnesses and scribes are given  after the first 10 lines.

We  list all personal names of  {\small JEN 208}. 

\noindent
{\footnotesize {\footnotesize JEN 208}
\begin{verbatim}
    line    name and kinship  information
    1   Ilu-ia，s. of H_{u}a-ma-at-ta-ar
    2   H_{u}a-ma-at-ta-ar，f. of I-lu-ia
    3   Pu-h_{u}i-s^{v}e-en-ni，f. of Te-h_{u}i-ip-til-la
    6   Im-bi-li-s^{v}u
    7   E-ni-ia
    8   Ilu-ia，s. of H_{u}a-ma-at-ta-ar
    8   Te-h_{u}i-ip-til-la，s. Pu-h_{u}i-s^{v}e-en-ni
   10   Te-h_{u}i-ip-til-la，s. Pu-h_{u}i-s^{v}e-en-ni
   11   Ilu-ia，s. of H_{u}a-ma-at-ta-ar
   13   Ilu-ia，s. of H_{u}a-ma-at-ta-ar
   13   Te-h_{u}i-ip-til-la，s. Pu-h_{u}i-s^{v}e-en-ni
   14   Ilu-ia，s. of H_{u}a-ma-at-ta-ar
   15   A-ta-a-a，f. of Ma^{n}r-^{d}is^{v}tar(U)
   16   It-h_{u}i-ip-s^{v}arri，s. of Te-h_{u}u-ia
   17   Ki-li-ip-s^{v}e-ri，s. of Na-as^{v}-w<i>
   17   Na-pu，f. of Ki-li-ip-[s^{v}]e-ri
\end{verbatim}
}

We make a database 
just  from NPN, which is 
  an index of about
ten thousands of individuals  (about 95 \%  of them are male)  who appear in cuneiform
 tablets  
excavated from the site of Nuzi.   

Several problem, however, should be solved when constructing a data base from 
NPN and identify the individuals.  

1) Many persons share the same name, 

2) Personal names might be written with different cuneiform characters,  

3) Some names might be overlapping, 

for example, name A can be a single name, but also included 
in two, three or more generation family trees.  These problems were avoided 
as much as possible by considering 
other information such as reference numbers and line numbers. 

We reconstruct  family trees from NPN. 
We apply the criterion that 
two persons with the same name are regarded as the same person,

(R1) if the both persons appear on the same line of the same document and regarding
them as the same person does not make any contradiction. 

\noindent 
or

(R2)  if the both persons appear at least  in three common documents and
 their relation is consistent.

\begin{figure} [t]
\special{epsfile=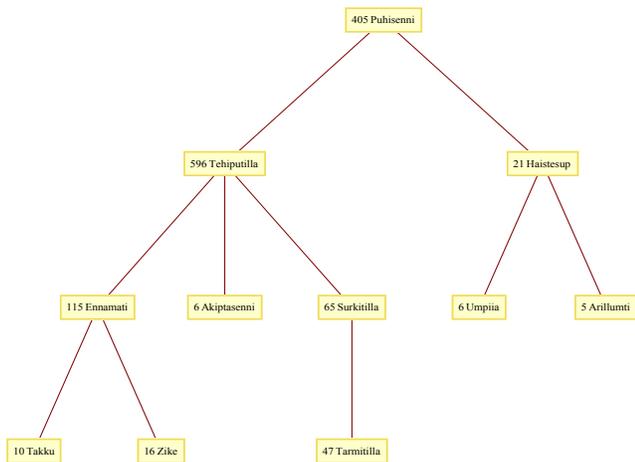 hscale=0.65  vscale=0.6}
\vspace{6cm}
\caption{Te\b{h}iptilla and  his  father, brother, son, and grandson.  The number of appearance 
of each name 
in documents is shown as 596Tehiptilla.  
The  birth year and death year of each name \cite{u},    in the bracket ()  is as, 
Pu\b{h}i\v{s}enni  (39.06-- 99.06),   
 Te\b{h}iptilla (59.06--114.06),  
 Hai\v{s}te\v{s}up  (63.27--114.06),  
Akipta\v{s}enni  (74.35--134.27),   
Ennamati (74.06--114.06),   
\v{S}urkitilla  (74.07--105.69),  
Arillumti  (78.27--135.74),   
Umpiia(78.27--114.06 ), 
Takku (89.06--139.78),  
Zike  (89.06--116.70),  
Tarmitilla (89.07--137.56). }\label{tehip}
\end{figure}

For example from the above (R1), we find 1) and 5) of Iluia in Example 2 is the same person. 
 After applying the two criterions  (R1) and (R2), 
we obtained a total of 10343 family trees.
We  
arrange sequentially the family trees obtained directly from the data, 
as $f_1, f_2, ..., f_{10343}$. 
Let us say the  family tree $A$ and the family tree $B$ are   consistent with each 
other, 
if  and only if at least  two names are common  in the $A$ and the $B$ and 
 there is no contradiction 
in their kinship relations for the names both in the $A$ and the $B$. 
Starting from  the initial set of the 10343 family trees,    
we apply  the  sequential 
 algorithm   recursively  to unify  mutually consistent  family trees.  
We continue  the procedure 
until there is  no family tree  to be unified \cite{iiut, umi} and get   family trees as  shown  
in Fig \ref{tehip}, whch is a part of 16 members obtained by our procedure.  
 The Te\b{h}iptilla family tree estimated by our sequential 
method,  using just NPN, is about the half size of the one  \cite{mai, mai2} , 
constructed 
by using  all possible information in contents of cuneiform tablets,  and does not coincide 
with 
 it in part.

We obtained 1 family tree with 6 generations 16 members, which is 
the largest family tree obtained , the Te\b{h}iptilla family tree given in part  
in Fig  \ref {tehip}, 
2 family trees with 5 generations, one is with  5 members and the other is with 11 members, 
3 family trees with 4 generations 4 members, and 
61 family trees with 3 generations. 
4096 family trees with 2 generations and family trees and 1987 single family trees 
and 24 family trees of 
one generation with size 2. 
Here we note that most of the 2 generation family trees are pairs of a  father and
his son, and a single family tree means a  person without any given  kinship in NPN.
The Te\b{h}iptilla family existed for 6 generations.  
If the two personal names appear in a tablet, 
we assume the two persons lived in the  same period. 
Based on this idea, we associate each individuals to one of the six 
generations of the  Te\b{h}iptilla family. 

Consider the family of  \b{H}amattar,  in Example 1, 2, of three generations. 
Ilu\b{i}a, the son of \b{H}amattar, is in the tablets JEN 208  and JEN 369.  
Ta\b{i}a, the grand son of \b{H}amattar, is   in just  JEN 369. 
Pu\b{h}i\v{s}enni Te\b{h}iptilla the son of Pu\b{h}i\v{s}enni and Enna-mati 
the son of Te\b{h}iptilla 
are in the tablets  JEN 208, JEN 369, and  JENu 414. Other personal names of the
 Te\b{h}iptilla 
family are not in the tablets. 
So let Pu\b{h}i\v{s}enni be the first generation, and Te\b{h}iptilla be the second generation. 
\b{H}amattar is approximately considered to be  in  the first generation, Ilu\b{i}a  in the second generation and 
Ta\b{i}a in the third generation.  So we can see   approximate correspondences of 
generation among family trees.

We observe that most of the names in NPN are in one  person family tree or 
the two person  family tree  \cite{umi}. 
They may be considered as ordinary  people in obscurity.  
They 
 appear in the cuneiform tablets 
only in the earlier period of the logistic growth   Fig \ref{logistic}.  After the contracts 
they  do not  appear in 
cuneiform tablets. 
 45\% of them share documents with  Te\b{h}iptilla family indicates that 
the Te\b{h}iptilla family 
occupied 
the high rank of the 
Nuzi society.  Through the false adoption contracts, ownership of land estates and other 
properties
 of them 
 were passed to the hands of Te\b{h}iptilla family.  Several other high ranked families were 
also observed. 
17 families with 
79 people shares documents with 70\%.  The results show the 
social 
hierarchy in the city of 
Nuzi  \cite{le}.

The  dynamical system, for the logistic growth, 
$$ \frac{d}{d~t} f(t)=\frac{1}{\beta}  (1-f(t))f(t),$$
has the solution 
$f(t)=1/(1+e^{-\frac{t-\mu}{\beta}})$. 
The system  finds applications in    biology, demography, economics, 
chemistry, sociology, political science, and many other fields.
 For example 
in population genetics the relative abundance  of a  advantageous mutant gene
 increases in a population 
until saturation $f(t)=1$.   
Assuming the above dynamical  system for  the ratio  
  $f(t)$ of properties  owned 
by powerful families as the Te\b{h}iptilla  family,   we  have the probability density of 
logistic distribution 
function with mean $\mu$ and scale parameter $\beta$,
 $$\frac{ e^{-\frac{t-\mu}{\beta} }}{\beta(1+ e^{-\frac{t-\mu}{\beta}})^2}.$$
 which seems to  explain the distribution of publication years of the cuneiform tablets
 Fig \ref{logistic}. 
 We can imagine the land owner class appeared through the contracts, as in JEN 208 of 
Example 1, during the period and after the 
 period the society changes  to another stage of social hierarchy  in which lands are owned by  
few  families.  
 
Now we show the outline of our  optimization method how we estimate 
the the publication year of each document, birth year and the  
death year of each individual,
consistently with NPN.
As we easily imagine NPN  alone is not enough to 
restore a detailed scenario of population dynamics of society. 
 The index just contains (partial) kinship information and 
about those people who appear in a document
simultaneously, that  is to say, were alive at the  time of contract. 
To fill this gap, we make use of a reasonable prior
information and employ a simulation and optimization approach.

First we give an initial lifespan to each person uniformly at random on the interval $[20,60]$
 independently.  
Based on this given initial set of lifespans of people, 
we formulate an optimization problem to find a configuration of 
birth year,  death year of individuals and publication year of cuneiform tablets 
consistent with the information from NPN.
Through solving the optimization problem, we obtain a configuration
which is closest to the initial lifespan
among all configurations consistent with the document information. 

Though the obtained configuration is different depending on the initial
lifespan given at random.  We find a common social structure among the obtained
configurations which would reflect what was going on in ancient Nuzi.

We formulate the optimization problem for unknown variables, 
 the birth year $b_i$ and the death year  $d_i$ of the person $i$, 
the id number of the father  $f_i$  and the mother  $m_i$  of the person $i$. and 
 the published year $P_k$ of the document $k$. 
The quadratic optimization problem to minimize
$\sum _i(d_i - b_i - \mu_i)^2$  finds a configuration, whose lifespans are closest to the
initial lifespans $\mu_i$, 
 in the least square sense, subject to  that

(I) $b_i \geq  b_{f_i }+ g_f$, ( a male becomes a father elder than $g_f$  years old 
for all person $i$ whose father appear in NPN), and  $d_{f_i}\geq b_i$,
 $b_i \geq b_{m_i} + g_m$  (a female becomes a mother elder
than $g_m$  years old, 
for all person $i$ whose mother appear in NPN),  
and $d_{m_i} \geq b_i$. 

(II)  $P_k \geq b_i + g_p, ~~d_i \geq P_k$ 
( a person appeared in a document  is supposed 
to be greater  than $g_p$ years old  
 for all  documents $k$ and person $i$ in the document $k$, and $i$ lived at the time of 
contract ),

(III)  $P_{906} = 100$   to fix  the origin of coordinate,  where 906 is the id for
 JEN 525.   
JEN 525
contains 52 different individual names.    
We make a set of names, each of  which is connected with directly or indirectly to one of the member through kinship relation 
or common contracts,  and  call it extended 
members of JEN 525. For each document we make extended members. JEN 525 has the largest extended 
members among all documents which includes Te\b{h}iptilla's family.

The above least square  problem was solved with a software
package NUOPT (Mathematical Systems Inc.).
The optimization result is  depending on the
initial set of lifespans. We conducted the  computation for Fig  \ref{logistic} 
assuming $g_f=15$, 
$g_m=20$, and $g_p=10$. 

The   result  of  
our reconstructing  family tree makes the number of variables  and constraints 
 smaller, as we  find 1) and 5) of Iluia in ex.2 is the same person. 
A total of 10060 persons appears
in our study of NPN.  We denote by $\mu_i$ the lifespan of the person $i$,  for  
$i = 1, . . . , 10060$. 
The number of documents is 1662.  
 
The convex quadratic program with 21782(= 10060 $\times$ 2 
(representing $b_i, ~~d_i$ ) + 1662
(representing $P_k$)) variables.  
We have  39357 constraints. The above  (I) makes 4241 $\times$ 2 constraints of them, 
  (II)  makes 
15437
 $\times$ 2 constraints of them,   
 and  
(III) makes 1 constraint of them.

 We carried out the optimization 10 times independently,  changing the  initial span  
given at random. 
The structure of the optimal solutions are more or less the
same regarding to the properties we reported here.

 \begin{acknowledgments}
This work is in part supported by Ministry of Education Science and Culture of Japan 
( Research Project 07207245, and 
 Research Project  09204245), and  ISM Cooperative Research (2011- ISMCRP -1022). 
Y. I. is supported in part 
by
US National Science Foundation Grant DMS 0443803 to Rockefeller University.
\end{acknowledgments}

\end{document}